\documentclass[11pt,twocolumn,tighten,twocolappendix]{aastex63}
\usepackage{hyperref}
\usepackage{color}
\usepackage{url}
\usepackage{natbib}
\usepackage{graphicx}
\usepackage{xfrac}
\usepackage{ulem}
\usepackage{CJK}
\usepackage{threeparttable}
\usepackage{multirow}
\usepackage{amsmath}
\begin{document}
\shorttitle{Environments of Giant Radio Galaxies}
 \shortauthors{T.W. Lan \& J. X. Prochaska}
\title{On the Environments of Giant Radio Galaxies} 
\begin{CJK*}{UTF8}{bsmi}
\author{Ting-Wen Lan 藍鼎文}
\affil{Department of Astronomy and Astrophysics, UCO/Lick Observatory, University of California, \\ 1156 High Street, Santa Cruz, CA 95064, USA}
\author{J. Xavier Prochaska}
\affil{Department of Astronomy and Astrophysics, UCO/Lick Observatory, University of California, \\ 1156 High Street, Santa Cruz, CA 95064, USA}
\affil{Kavli IPMU, the University of Tokyo (WPI), Kashiwa 277-8583, Japan}

\begin{abstract}
We test the hypothesis that environments play a key role in enabling the growth of enormous radio structures spanning more than 700 kpc, an extreme population of radio galaxies called giant radio galaxies (GRGs).
To achieve this, we explore (1) the relationships between the occurrence of GRGs and the surface number density of surrounding galaxies, including satellite galaxies and galaxies from neighboring halos, as well as (2) the GRG locations towards large-scale structures. The analysis is done by making use of a homogeneous sample of 110 GRGs detected from the LOFAR Two-metre Sky Survey in combination with photometric galaxies from the DESI Legacy Imaging Surveys and a large-scale filament catalog from the Sloan Digital Sky Survey.
Our results show that the properties of galaxies around GRGs are similar with that around the two control samples, consisting of galaxies with optical colors and luminosity matched to the properties of the GRG host galaxies. Additionally, the properties of surrounding galaxies depend on neither their relative positions to the radio jet/lobe structures nor the sizes of GRGs. We also find that the locations of GRGs and the control samples with respect to the nearby large-scale structures are consistent with each other. 
These results demonstrate that there is no correlation between the GRG properties and their environments traced by stars, indicating that external galaxy environments are not the primary cause of the large sizes of the radio structures.
Finally, regarding radio feedback, we show that the fraction of blue satellites does not correlate with the GRG properties, suggesting that the current epoch of radio jets have minimal influence on the nature of their surrounding galaxies.
\end{abstract}

\section{Introduction}
Over the past two decades, feedback from supermassive black holes has been considered as a fundamental mechanism in galaxy formation theory in order to reproduce the observed properties of galaxies \citep[e.g.,][]{Benson2003, Schaye2015, Dave2019, Nelson2019}. Supermassive black holes are expected to remove a substantial mass of gas from galaxies and thereby
quench star formation. In addition, supermassive black holes are expected to maintain the heat content of gas around quenched galaxies to prevent gas cooling for further star-formation activity \citep[e.g.,][]{Silk1998,Bower2008,Alexander2012,McNamara2012,Chen2019,Hardcastle2019}. 
%This feedback mechanisms consist of different modes.

\begin{figure}
\center
\includegraphics[width=0.47\textwidth]{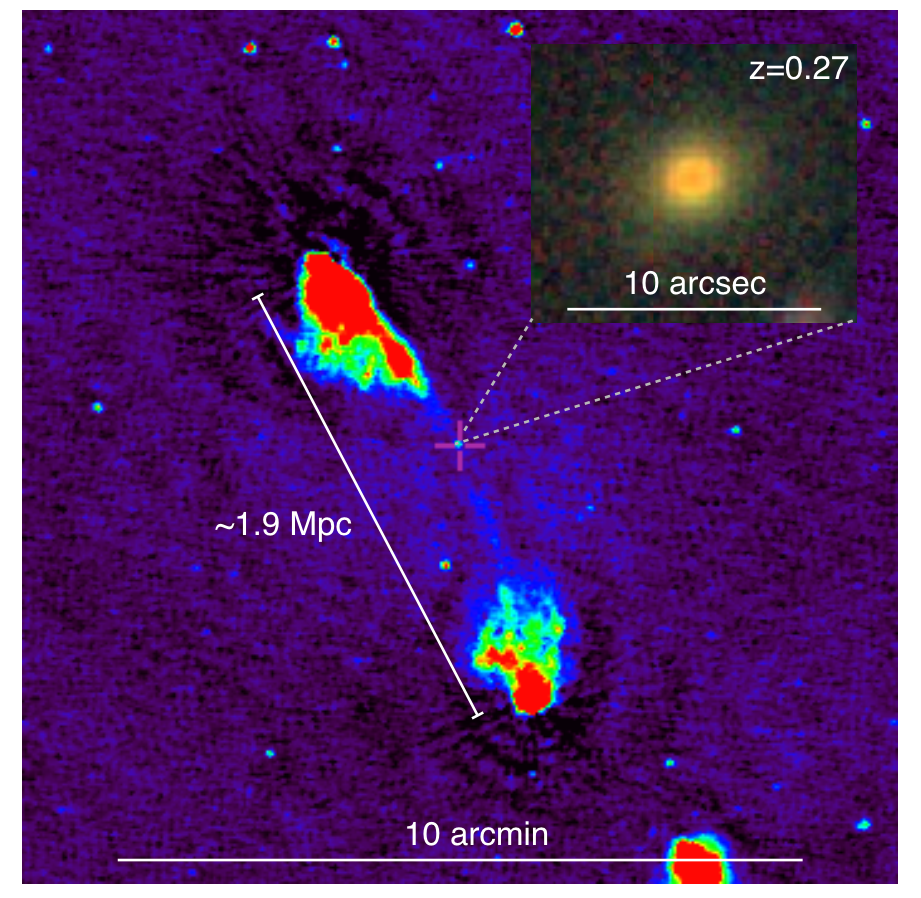}
\caption{Radio image of a giant radio galaxy at $z\sim0.27$ from the LOFAR Two-metre Sky Survey. The projected linear distance between the two radio lobes is approximately 1.9 Mpc. The optical image of the host galaxy (drawn from the DESI Legacy Imaging Surveys) is shown on the upper right corner.}

\label{fig:example_GRG}
\end{figure}

One of the observed form of black hole feedback is extended radio jets and lobes emerging from the centers of galaxies \citep[e.g.,][]{Worrall2008, best2012, Fabian2012, McNamara2012}. Since their discovery, radio galaxies have been classified into different types according to their morphological structure of radio emission \citep[e.g.,][]{Fanaroff1974,Hardcastle2020}. 
A small fraction of radio galaxies ($\sim 5\%$)
possess extremely extended radio-emission extending far beyond the size of dark matter halos and are referred to as giant radio galaxies (GRGs), 
\citep[e.g.,][]{Willis1974,Saripalli1986, Mack1998,Ishwara1999}. 
These GRGs are roughly defined as radio galaxies with linear projected distances between two radio lobes of greater than 700 kpc (See Figure~\ref{fig:example_GRG} as an example).
Given the extreme properties of GRGs, it is surmised that 
understanding the nature and origin of GRGs will illuminate 
the physical mechanisms of black hole feedback on galaxy formation. 

Previous studies have suggested two main scenarios that allow radio galaxies to grow to such large scales: (1) They can be produced by either powerful engines within normal timescales ($\sim10-100$ Myr) \citep[e.g.,][]{Gopal1989} or modest energy with prolonged timescales or multiple phases of activity \citep[e.g.,][]{Subrahmanyan1996,Hardcastle2019,Bruni2020};
(2) They tend to live in relatively low density environments which allows the radio lobes to grow to very large scales \citep[e.g.,][]{Mack1998,Malarecki2015}. 
While previous studies have explored both these scenarios, the origin(s) of GRGs are still not well understood.  This is primarily due to their rare and
difficult detection lending to small sample sizes.
Therefore, previous results are based on the properties of only a handful of GRGs. 

In the past few years, large radio surveys have increased the sample size of GRGs. 
For example, \citet{Kuzmicz2018} compile a sample of $\sim 300$ GRGs from all the previous observations and \citet{Dabhade2017} and \citet{Dabhade2020a}  detect $\sim200$ new GRGs from the NVSS dataset \citep[][]{Condon1998}. Additionally, based on the DR1 data of the LOFAR Two-metre Sky Survey \citep[LoTSS,][]{Shimwell2019}, \citet{Dabhade2020} compile a sample of $\sim 200$ GRGs covering a wide range of redshift. These new GRG samples offer the 
opportunity to better characterize the properties of GRGs and their surrounding environments.  

In this work, we will test the hypothesis that GRGs tend to live in low density environments by measuring the environments of GRGs as a function of redshift.
To this end, we will make use of a GRG catalog compiled from the LoTSS and data from large optical surveys, including the imaging Surveys for the Dark Energy Spectroscopic Instrument (DESI)  \citep[][]{Dey2019} and the Sloan Digital Sky Survey \citep[][]{York2000}. The results will provide strict constraints on this hypothesis and thereby shed new light on the driving factors for the enormous sizes of GRGs.  
The structure of the paper is as follows. Our data analysis is described in Section 2. We show our results and discuss their implications in Section 3 and Section 4. We summarize in Section 5. Throughout the paper we adopt a flat $\Lambda$CDM cosmology with $h=0.7$ and 
$\Omega_{\rm M}=0.3$. When referring to distances, we use physical distances.

%-------------------------------------
%--------------
%-------------------------------------
\begin{figure*}
\center
\includegraphics[width=0.85\textwidth]{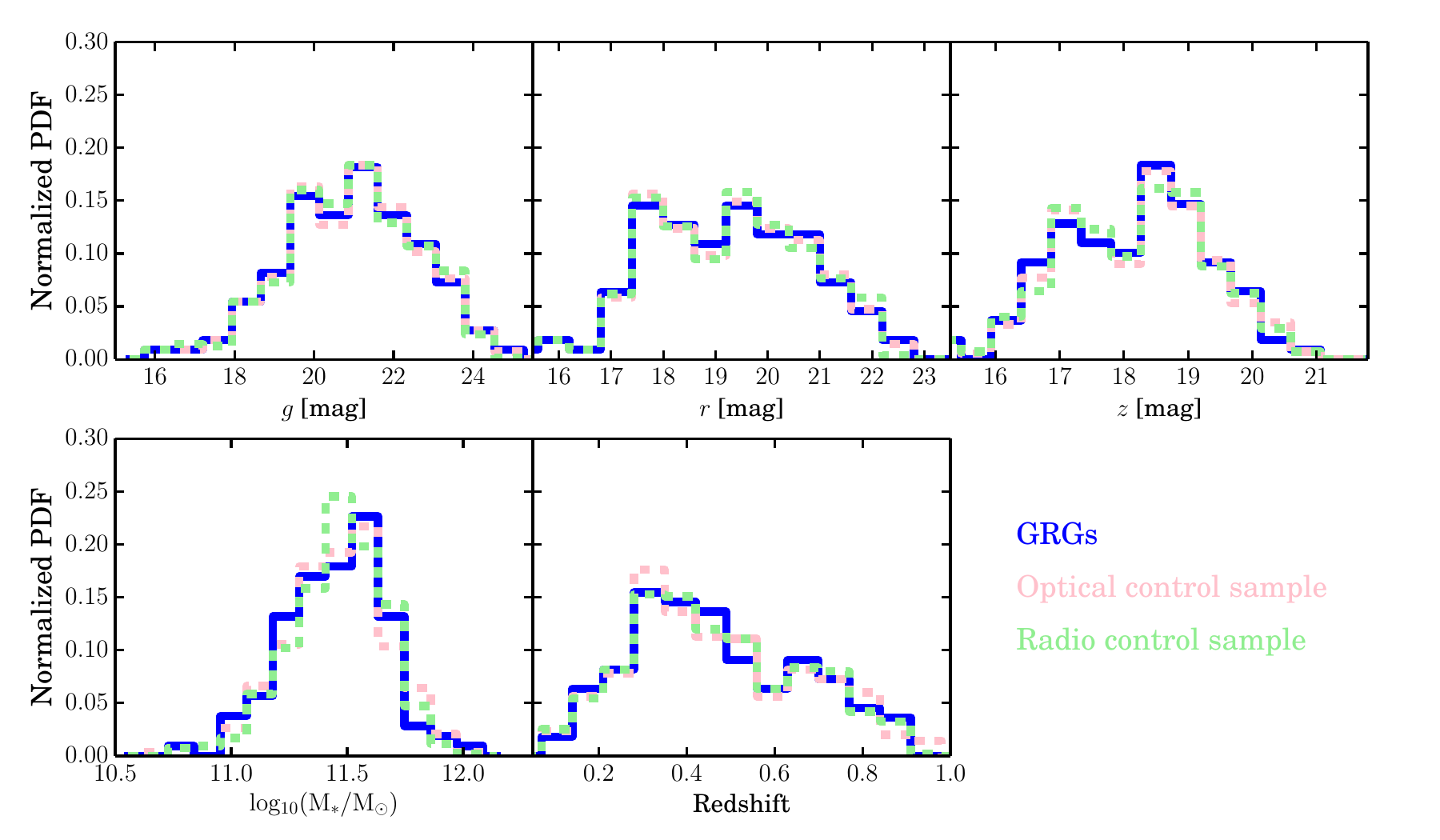}
\caption{Magnitude, stellar mass, and  redshift distributions of the GRG host galaxies (blue) and the radio (green) and optical (red) control samples.}
\label{fig:control}
\end{figure*}
%--------------
\section{Data Analysis}
To investigate the local and large-scale environments of GRGs, we apply statistical cross-correlation analyses to a large GRG catalog, a photometric galaxy catalog and a large-scale filament catalog. By doing so, we are able to quantify the local and large-scale environments of GRGs and compare the measurements with that of the control samples. 
Our approach, therefore, is to isolate whether the presence of 
bright and highly extended radio emission is correlated with the 
underlying  galactic environment.
In what follows, we describe the data sets and methods we use.

\subsection{Datasets}
\begin{itemize}
    \item \textbf{Giant radio galaxy catalog}: We make use of a large, homogeneous giant radio galaxy catalog compiled by \citet{Dabhade2020}. This sample is obtained from the 1st data release\footnote{\url{https://lofar-surveys.org/releases.html}} of the LOFAR Two-metre Sky Survey \citep[LoTSS, ][]{Shimwell2019, Williams2019}, consisting of 239 GRGs from $z\sim0$ up to $z\sim1$. These GRGs are identified from the LoTSS pipeline assisted by visual inspections. 
    
    To construct control samples as described in Section 2.2, we select GRGs that have 
    (1) spectroscopic redshifts of their host galaxies identified in the 14th data release of the Sloan Digital Sky Survey \citep[SDSS,][]{York2000, Abolfathidr14} and (2) imaging observations covered by the DESI Legacy Imaging Surveys \citep{Dey2019}. We exclude 40 SDSS sources identified as quasars.
    This yields a sample of 110~GRGs for our analysis. As an example, Figure~\ref{fig:example_GRG} shows the LOFAR radio image of a GRG and the corresponding optical image (g, r, and z bands) of the host galaxy. 
    
    \item \textbf{Photometric galaxy catalog}: To investigate the local galaxy environments of GRGs, we use the DR8 photometric galaxy catalog\footnote{\url{https://www.legacysurvey.org/dr8/}} of the DESI Legacy Imaging Surveys \citep[the DESI Legacy Surveys,][]{Dey2019}. The catalog contains $\sim 10^{9}$ galaxies over $\sim 14,000\, \rm deg^{2}$ of the sky with photometric properties of $g$, $r$, $z$ and WISE channel 1, 2, 3, 4 bands \citep{Wright2010} characterized by {\it the Tractor} algorithm \citep{Lang2016}. The limiting magnitudes of the DESI Legacy Surveys are about $24$, $23.4$ and $22.5$ for $g$, $r$, and $z$ bands respectively. These depths are sufficient for our purposes of detecting $L^{*}$ galaxies up to $z\sim1$.
    
    \item \textbf{Large-scale filament catalog}: For exploring the locations of GRGs with respect to the large-scale structure, we use a filament catalog\footnote{\url{https://sites.google.com/site/yenchicr/home}} from \citet{Chen2016catalog}. The filamentary structures are identified based on a statistical algorithm for detecting density ridges \citep{Chen2015method} applied to the SDSS spectroscopic galaxy sample. The filament catalog includes random points sampled from the filaments as a function of redshift up to redshift 0.7 with 0.05 redshift intervals.
\end{itemize}

\subsection{Properties of the GRG host galaxies}
\label{sec:GRG}
We summarize the properties of the GRG host galaxies in Figure~\ref{fig:control}, which shows the magnitude, stellar mass, and redshift distributions of the host galaxies.  
The stellar masses of the GRG host galaxies are obtained via spectral energy distribution (SED) fittings with the {\it CIGALE package} \citep{Boquien2019}. We adopt the simple stellar population from \citet{Bruzual2003} with Chabrier initial mass function \citep{Chabrier2003} and delayed-exponential star-formation history. The best-fit stellar mass is constrained by the observed flux in $g$, $r$, $z$, WISE 1 and WISE 2 bands.

Similar to the host galaxies of typical radio-AGN galaxies \citep[e.g.,][]{Best2005, Lin2010}, the GRG host galaxies are massive red galaxies with $M_{*}>10^{11}\, M_{\odot}$. The corresponding typical dark matter halo mass is therefore
a few times of $10^{13}\, M_{\odot}$ and the virial radius is approximately 500-600 kpc \citep[e.g.,][]{Tinker2016}. About 10\% of GRGs are located in dark matter halos with even higher masses ($>10^{14}\, M_{\odot}$), i.e.\  galaxy cluster environments \citep[e.g.,][]{Dabhade2020}.
%-------------------------------------
\begin{figure*}
\center
\includegraphics[width=0.85\textwidth]{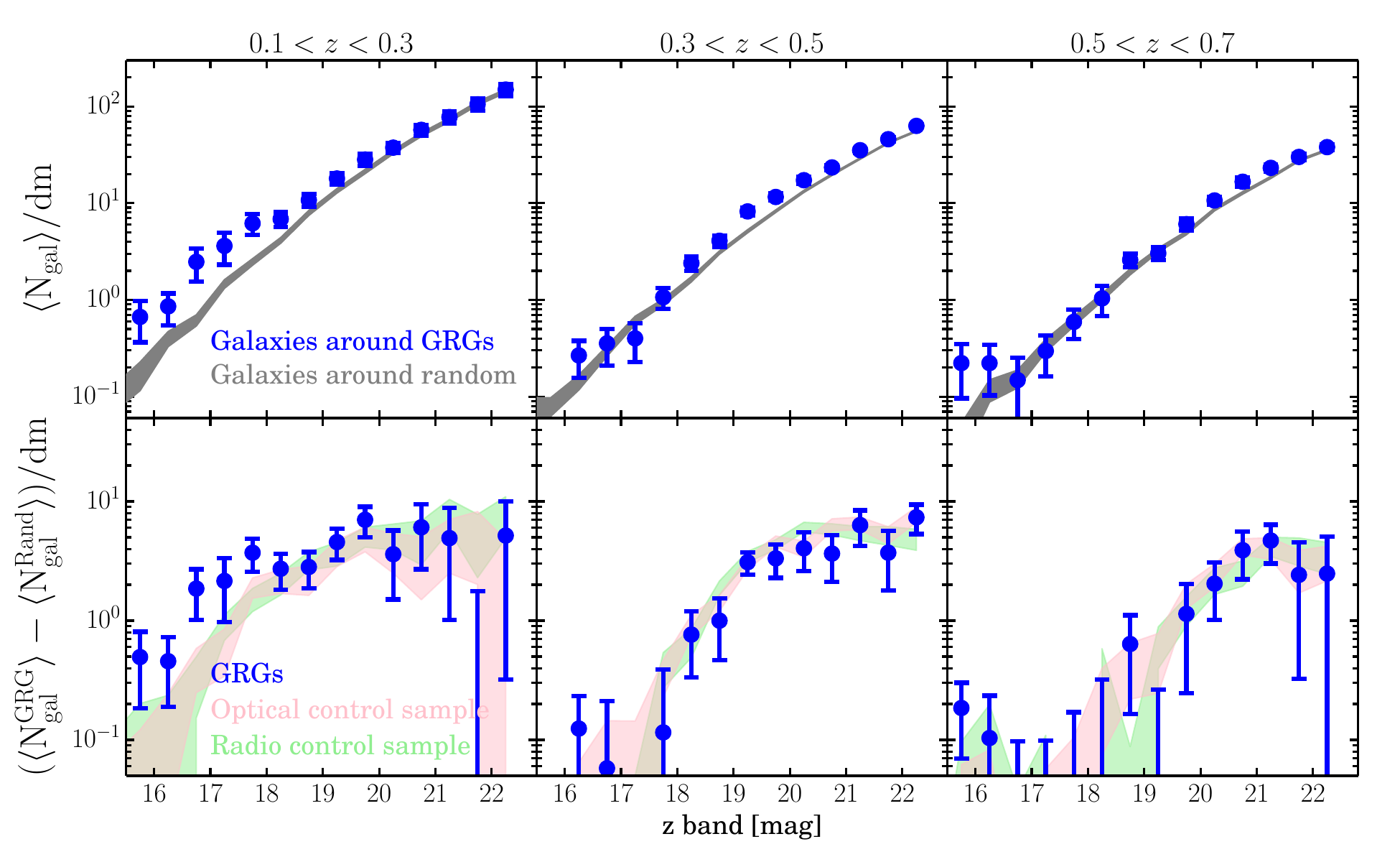}
\caption{\emph{Top:} Average number of galaxies around GRGs (blue) and around random positions (grey) within 600 kpc as a function of $z$-band magnitude and redshift. The difference between the two shows the contribution from satellite galaxies associated with GRGs as shown by blue data points in the bottom panel. 
\emph{Bottom:} Magnitude distribution of satellite galaxies around GRGs. The red and green shaded regions show the same measurements around the optical and radio control samples respectively. The magnitude distributions of satellite galaxies around GRGs are consistent with the distributions around the two control samples.}
\label{fig:magnitude_illustration}
\end{figure*}

%--------------
\subsection{Control samples}
To evaluate whether or not GRGs tend to live in special environments, we construct two control samples that have similar optical photometric properties and redshifts as the GRG host galaxies. One sample is selected to have radio-emission detected in LoTSS with integrated flux greater than 0.2 mJy and the other is primarily selected to have no LOFAR detection.
When combined, we can examine whether the presence of {\it any} radio emission is correlated to galaxy environments. In the following, we describe the two control samples in detail:
\begin{enumerate}
    \item \textbf{Radio control sample:} To construct the radio control sample, we first cross-match SDSS DR14 spectroscopic galaxies {\it with radio sources} detected in LoTSS \citep{Williams2019, Duncan2019}, excluding the GRGs in the sample. 
    This radio-optical sample consists of approximately 30,000 objects. 
    For each GRG, we then select five galaxies from this radio-detected SDSS galaxy sample as the corresponding radio control galaxies. These control galaxies are selected with redshifts, $g$, $r$, $z$ band magnitudes, and morphology type matched to that of the host galaxy of the GRG.
    Specifically, these lie within 0.025 difference in redshift 
    and 0.05 magnitude difference in each band. If fewer than 5 galaxies 
    satisfied this selection criteria, we increase the redshift range by $20\%$ and magnitude difference by 0.05 iteratively. 
    \item \textbf{Optical control sample:} To construct the optical control sample,  
    we make use of SDSS DR14 spectroscopic galaxies located between $\rm 162<RA<235$ deg and $\rm 46<DEC<56.5$ deg, approximately the 
    sky region covered by the LoTSS DR1. We then exclude all of the SDSS galaxies detected in the DR1 LoTSS catalog. This process reduces the sample size by $\sim 10\%$. This optical sample consists of about 240,000 objects. For each GRG, we select five galaxies from this optical SDSS galaxy sample that satisfy the same criteria listed above for the radio control sample. We note that about $10\%$ of the optical selected region is not covered by the DR1 LoTSS. Therefore, $10\%$ of galaxies in the optical control sample are selected without radio information. Nevertheless, with the LoTSS radio detection rate of SDSS sources $(\sim 10\%)$, the contamination rate of radio emission in this optical control sample is expected to be nearly negligible (i.e.\ $\sim1\%$).
\end{enumerate}
By performing the same analysis to the GRG sample and the two control samples, we will compare the galaxy environment properties as a function of the extension of radio emission with fixed optical properties of host galaxies.
Figure~\ref{fig:control} shows the magnitude, stellar mass, and redshift distributions of the GRG host galaxies, the optical control sample and the radio control sample in blue, red, and green respectively. 
The stellar masses of galaxies in the control samples are obtained via SED fittings as described in Section~\ref{sec:GRG}.

\subsection{Methods}
\label{sec:analysis}
We focus on two statistical measurements to characterize the environments of GRGs. The first one is the surface number density of surrounding galaxies associated with GRGs obtained by statistically subtracting the number density of coincident background and foreground galaxies. We perform the measurements from 20 kpc to 2 Mpc in projected distances, which include satellite galaxies bounded within the host halos of GRGs (the so-called ``1-halo term'') and galaxies from neighboring halos (2-halo term).
Given that the properties of satellites are the primary concern of the analysis, we will refer to these measurements as the distributions of satellites for simplicity.

The second measurement is the distances between GRGs and the nearest filamentary structures, which are identified from the SDSS spectroscopic galaxies. The typical scale involved is $\sim10$ Mpc, several times larger than the satellite measurements. We use this GRG-filament distance measurement to infer the location of GRGs in the context of large-scale structure. 
In the following, we will describe the methods we apply for these two measurements respectively.
\subsubsection{Cross-correlation method for estimating the number of satellite galaxies}
To obtain the number density of satellite galaxies, we cross-correlate the GRG sample with the photometric galaxies detected from the DESI Legacy Surveys. 
For each GRG, we search and count the number of the surrounding photometric galaxies within a given (physical) impact parameter. The impact parameters are estimated by assuming that all the surrounding galaxies are at the same redshift of the GRG. The number of photometric galaxies around a given GRG, $\rm N_{gal}^{GRG}$, consists of two components: (1) satellite galaxies truly (physically) associated with the GRG, $\rm N_{sat}^{GRG}$, and (2) coincident background and foreground galaxies located at different redshifts, $\rm N_{coincident}$,
\begin{equation}
    \rm N_{gal}^{GRG} = N_{sat}^{GRG} + N_{coincident}.
\end{equation}
For each GRG, we estimate the average number of coincident background and foreground galaxies, $\rm \langle N_{coincident} \rangle$, based on the number of galaxies around 10 random positions in the sky footprint between $\rm 162<RA<228$ deg and $\rm 46<DEC<56.5$ deg. By subtracting $\rm \langle N_{coincident} \rangle$ from $\rm N_{gal}^{GRG}$, we can obtain the number of satellite galaxies associated with the GRG, $\rm N_{sat}^{GRG}$, without any spectroscopic redshift information of the photometric galaxies \citep[e.g.,][]{Lan2016}. Finally, we calculate the average number of satellite galaxies associated with the full set of GRGs to reduce Poisson uncertainty which dominates the number counts of individual sightlines, 
\begin{equation}
    \rm \langle N_{sat}^{GRG} \rangle = \frac{\sum\limits_{i=1}^{N_{GRG}} N_{sat}^{GRG,i}}{N_{GRG}}, 
\end{equation}
where $\rm N_{GRG}$ is the number of GRGs used in the analysis. 

\begin{figure*}
\center
\includegraphics[width=0.9\textwidth]{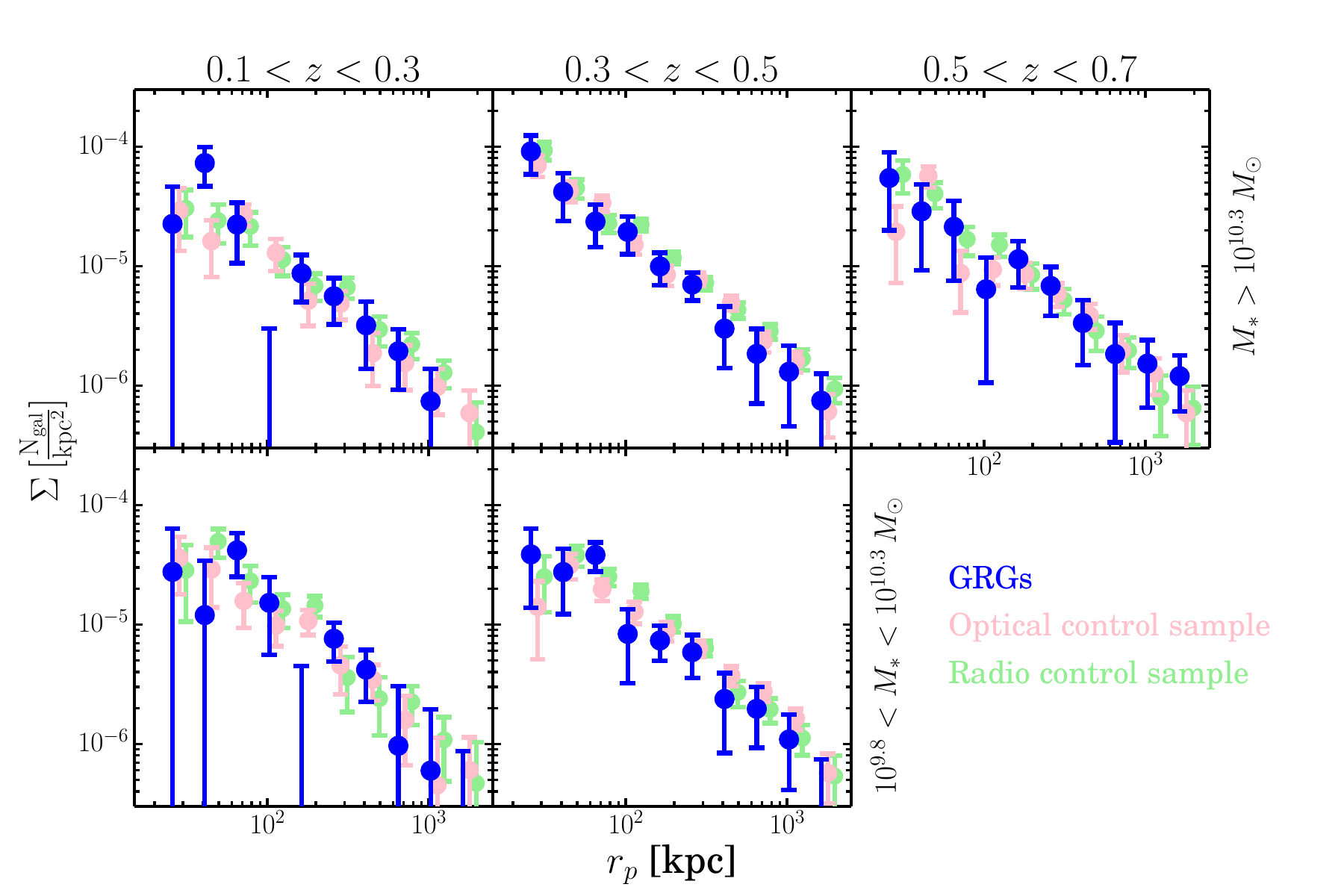}
\caption{Radial distributions of satellite galaxies around GRGs (blue), the optical control sample (red), and the radio control sample (green) as a function of redshift and stellar mass. The y-axis shows the surface number density of galaxies and the x-axis the impact parameters. The redshifts of the host galaxies increase from the left to the right panels. The upper and lower panels show the distribution for satellites with $M_{*}>10^{10.3} \, M_{\odot}$ and $10^{9.8}<M_{*}<10^{10.3} \, M_{\odot}$ respectively. 
The satellite radial distributions around GRGs and the two control samples are consistent with each other, indicating that GRGs do not preferentially live in lower density environments.}

\label{fig:radial_distribution}
\end{figure*}
%--------------

Figure~\ref{fig:magnitude_illustration} illustrates the method. In the upper panel, we show the average number of galaxies around GRGs (blue) and random positions (grey) within 600 kpc as a function of $z$-band magnitude and redshift. The lower panel shows the differences between the number counts around GRGs and around random positions. As can be seen, there are more galaxies around GRGs than around random positions. The difference results from the presence of satellite galaxies around GRGs. Additionally, the $z$-band magnitude distribution of these satellite galaxies becomes fainter towards higher redshifts. This trend confirms that the signal is only from satellite galaxies associated with GRGs. We find the number of satellites around GRGs within 600 kpc is similar to the number of satellites in dark matter halos with $\sim10^{13.5}-10^{13.9} M_{\odot}$ as shown in \citet{Lan2016}, similar to the halo masses inferred from the stellar mass of the host galaxies.
We note that this method has been applied to many other research areas, such as measuring the conditional luminosity functions of groups and clusters \citep[e.g.,][]{Lin2004, Lan2016, Tinker2019}, detecting the splash-back radius of dark matter halos \citep[e.g.,][]{More2016}, and probing galaxies associated with absorption line systems \citep[e.g.,][]{Lan2014, Lan2020}.

%-------------------------------------
\begin{figure*}
\center
\includegraphics[width=0.8\textwidth]{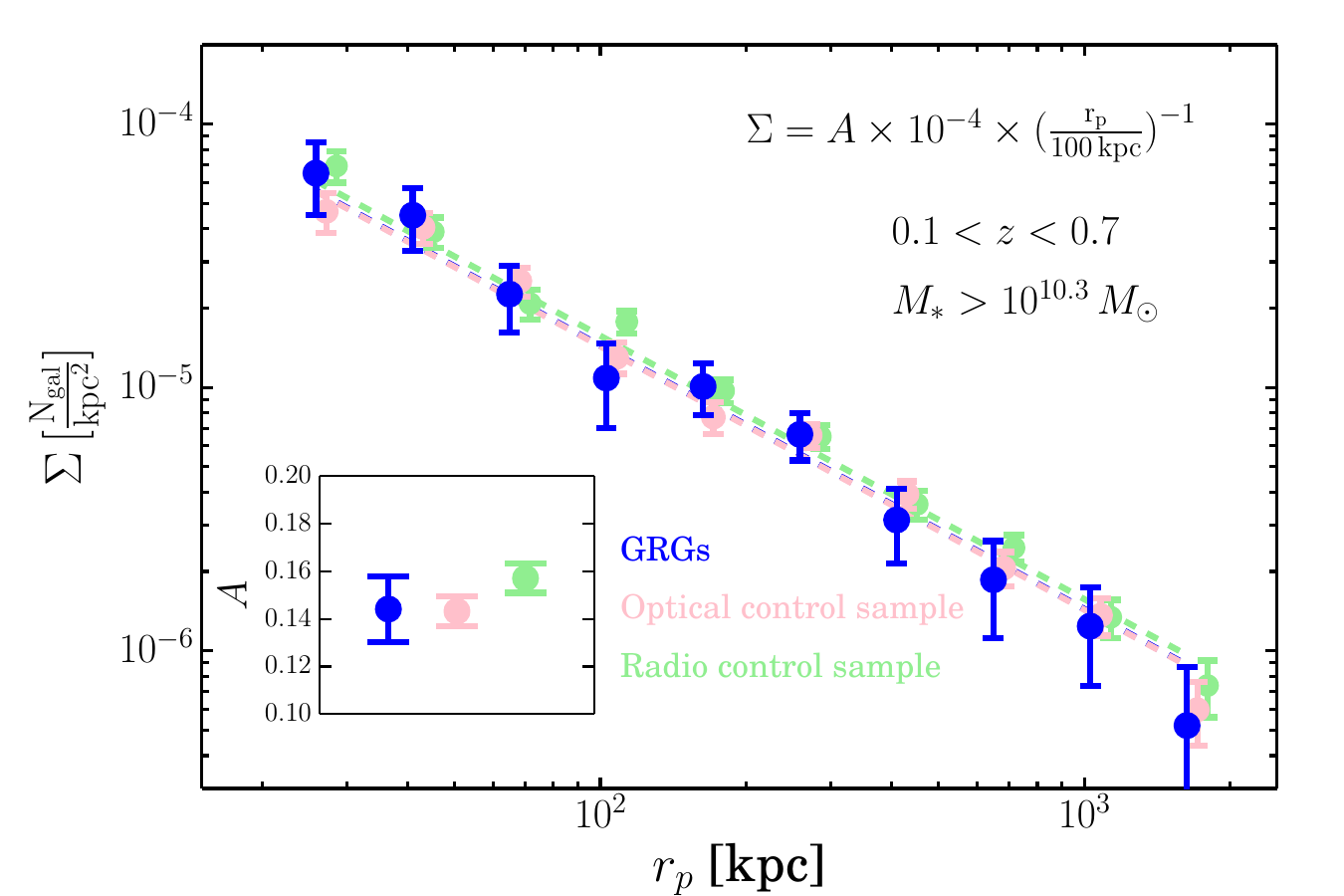}
\caption{Radial distributions of satellite galaxies with $M_{*}>10^{10.3} \, M_{\odot}$ around GRGs (blue) and the optical (red) and radio (green) control samples at redshift $0.1<z<0.7$. The inset shows the best-fit surface number density at 100 kpc for the three samples. The best-fit value for GRGs is consistent with the optical and radio control samples, indicating that the local galaxy environments of GRGs is mostly driven by the optical properties of the host galaxies.}
\label{fig:satellte_combine}
\end{figure*}
%--------------

In the lower panel of Figure~\ref{fig:magnitude_illustration}, we also show the magnitude distributions of satellite galaxies around the optical and radio control samples with the shaded red regions and green regions respectively. As can be seen, the satellite magnitude distributions of the two control samples are consistent with that of the GRG sample. We will make quantitative comparisons in Section~\ref{sec:results}. We note that we estimate the uncertainty by bootstrapping the sample 500 times. 

This method can be applied to explore any properties of galaxies associated with GRGs. Instead of the observed brightness of galaxies, we will explore the properties of satellite galaxies as a function of stellar mass. 
This will enable an examination of the full sample across redshift. To obtain the stellar mass of each galaxy, we use the {\it CIGALE package} \citep{Boquien2019} assuming that each photometric galaxy is at the same redshift of the corresponding GRG. We adopt the same approach as described in Section~\ref{sec:GRG}, with
the best-fit stellar masses constrained by the observed fluxes in the $g$, $r$, $z$, WISE 1 and WISE 2 bands. Note that
in our analysis, we focus on the stellar mass of galaxies in a relative sense. Therefore, the absolute systematic uncertainty of stellar mass introduced by the adopted assumptions of stellar population modeling do not affect our results. 
%-------------------------------------

\subsubsection{Distances between GRGs and the nearest filaments}
To better understand the large-scale structure environments in which GRGs reside, we measure the distance between each GRG and the nearest filament structure from the filament catalog.
The filaments are identified based on SDSS spectroscopic galaxies within the same redshift bin with a 0.05 redshift interval.
Therefore, for each GRG, we extract filaments with the corresponding redshift bin and calculate the projected distance between the GRG and the nearest filament structure.
We perform the same calculation for the control samples and compare the results between the three samples. The results will be described in Section~\ref{sec:results}.

%\newpage
\section{Results}
In this section, we will first present the distribution of satellite galaxies around GRGs and their control samples. We will also show the number of satellite galaxies around GRGs as a function of the properties of GRGs. Finally, we will present the environments in which GRGs reside in the context of large-scale structure. 
\label{sec:results}
\subsection{Radial distributions of satellite galaxies around GRGs and around the control samples}
Figure~\ref{fig:radial_distribution} shows the surface number density of satellite galaxies 
$\Sigma (r_{p}) = \rm \langle N_{sat}^{GRG}\rangle /area$, around GRGs and the control samples as a function of impact parameter $(r_{p})$, redshift and stellar mass. 
Here we probe the impact parameters from $20$ kpc to $2$ Mpc, which includes satellite galaxies within the virial radius ($\sim600$ kpc) of the GRG halos and galaxies in neighboring halos (the two-halo term).  
As can be seen, the radial distributions of satellite galaxies around GRGs and the control samples are consistent with each other across redshift and stellar mass.
For the rest of the paper, we focus on the distribution of satellites with $M_{*}>10^{10.3}\, M_{\odot}$, a stellar mass range that is complete to redshift 0.7 and is an order of magnitude less than the stellar mass of the GRG host galaxies (Figure~\ref{fig:control}).

In Figure~\ref{fig:satellte_combine}, we show the radial distributions of satellites with $M_{*}>10^{10.3}\, M_{\odot}$ around GRGs and the two control samples with $0.1<z<0.7$. We fit the satellite distributions with 
\begin{equation}
    \Sigma(r_p) = A\times10^{-4} \bigg(\frac{r_{p}}{100 \, \rm kpc}\bigg)^{-1},
\end{equation}
where $A$ is the surface number density of satellite galaxies at 100 kpc. We fix the power index as -1 given that the best-fit values are all consistent with -1 when the parameter is free. The best-fit $A$ parameter values are shown in the inset of Fig.~\ref{fig:satellte_combine}. 
As can be seen, the satellite distribution around GRGs is consistent with the distributions around the two control samples from 20 kpc to 2 Mpc.
This result demonstrates that the local galaxy environments of GRGs mostly depend on the optical properties of the host galaxies. In other words, GRGs do not reside in local environments significantly deviating from the environments of their optical and radio control samples. We note that this is observed at scales spanning from the inner region of the halos ($\sim 100 $ kpc) where the radio activity could potentially influence the surrounding galaxies to the two-halo region ($\sim 1$ Mpc) which primarily reflects the mass of dark matter halos hosting GRGs.

We emphasize that we can also interpret our results in the context of feedback. These measurements demonstrate that the presence of radio feedback does not affect the overall abundance of galaxies in the halos. In Section~\ref{sec:quenching}, we perform a more detail analysis measuring the fraction of blue galaxies. By doing so, we explore the possible effect of radio feedback on quenching star-formation in satellite galaxies. 

We note that there are slightly more satellite galaxies ($\sim10-20\%$) around the radio control sample than around the optical control sample as shown in the inset of Fig.~\ref{fig:satellte_combine}. 
%This trend is observed from 20 kpc to 2 Mpc. 
This result is similar to previous results showing that on average, radio-AGNs have more satellites than their control radio-quiet galaxies \citep[e.g.,][]{Mandelbaum2009, Donoso2010, Pace2014}. This might imply that with fixed stellar mass of the host galaxies, radio-AGNs tend to live in halos with higher masses than radio-quiet galaxies. Exploring this effect further is beyond the scope of this paper.

%-------------------------------------
\begin{figure*}
\center
\includegraphics[width=1\textwidth]{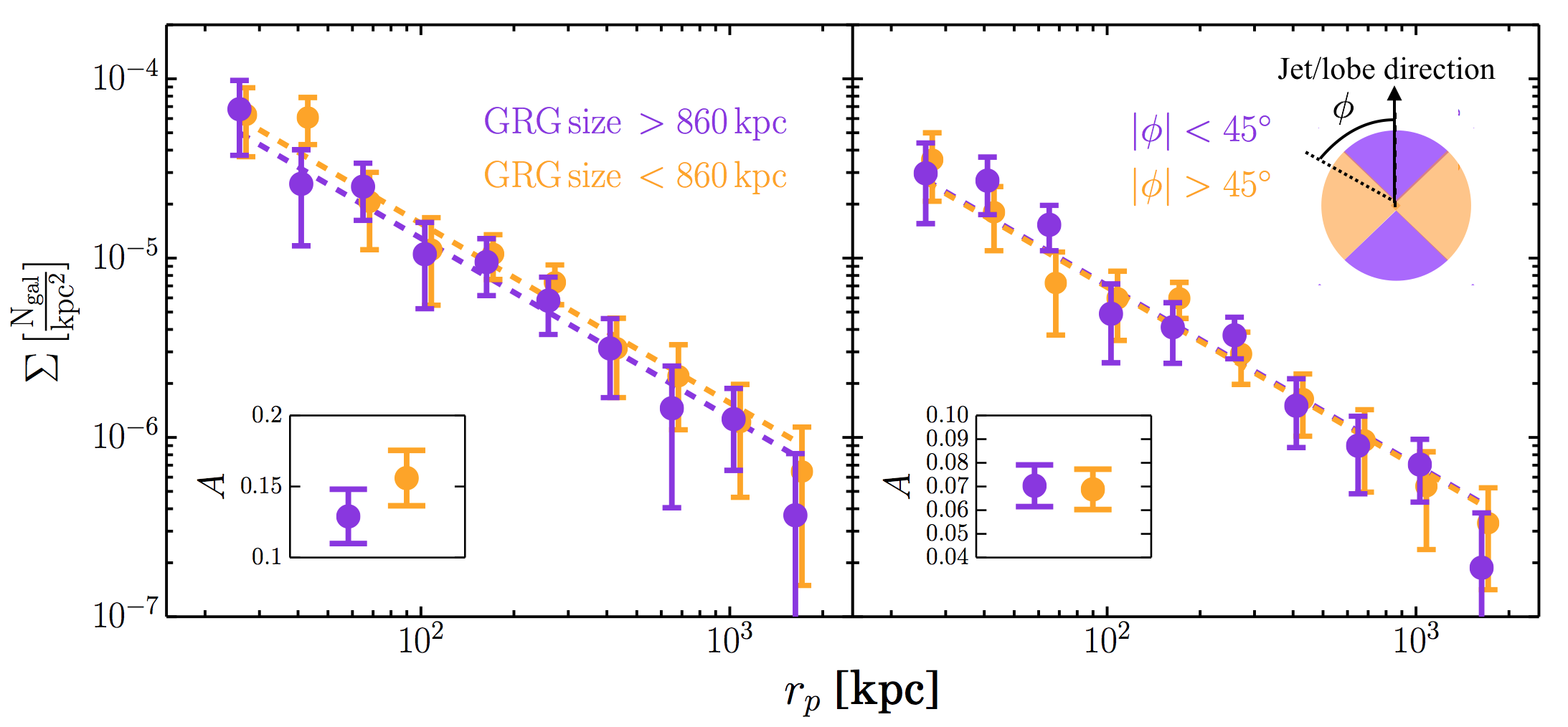}
\caption{Satellite distribution as a function of GRG properties. The distribution includes satellite galaxies with $M_{*}>10^{10.3} \, M_{\odot}$ and GRGs at $0.1<z<0.7$. \emph{Left:} Satellite distribution as a function of GRG size. The purple and orange data points show the satellite distributions around GRGs with sizes larger than 860 kpc, and smaller than 860 kpc respectively. \emph{Right:} Satellite distribution as a function of azimuthal angle. 
The purple and orange data points show the satellite distributions around GRGs towards and perpendicular to the radio lobe directions respectively. The insets in the panels show the best-fit A parameter values.}

\label{fig:size_az}
\end{figure*}
%--------------

\subsection{Radial distribution of satellites as a function of GRG properties} 
We now explore the radial distribution of satellite galaxies as a function of GRG size, i.e., the projected distance between two radio lobes. We separate the GRG sample into two equal number sub-samples according to their sizes.  The corresponding median value of the GRG sizes is about 860 kpc.
The left panel of Figure~\ref{fig:size_az} shows the distribution of satellites around GRGs with sizes larger than 860 kpc and smaller than 860 kpc. We find that the distribution of satellites does not correlate with the sizes of GRGs. In other words, the local galaxy environments do not play a role in the sizes of GRGs. 

%-------------------------------------
\begin{figure*}
\center
\includegraphics[width=1\textwidth]{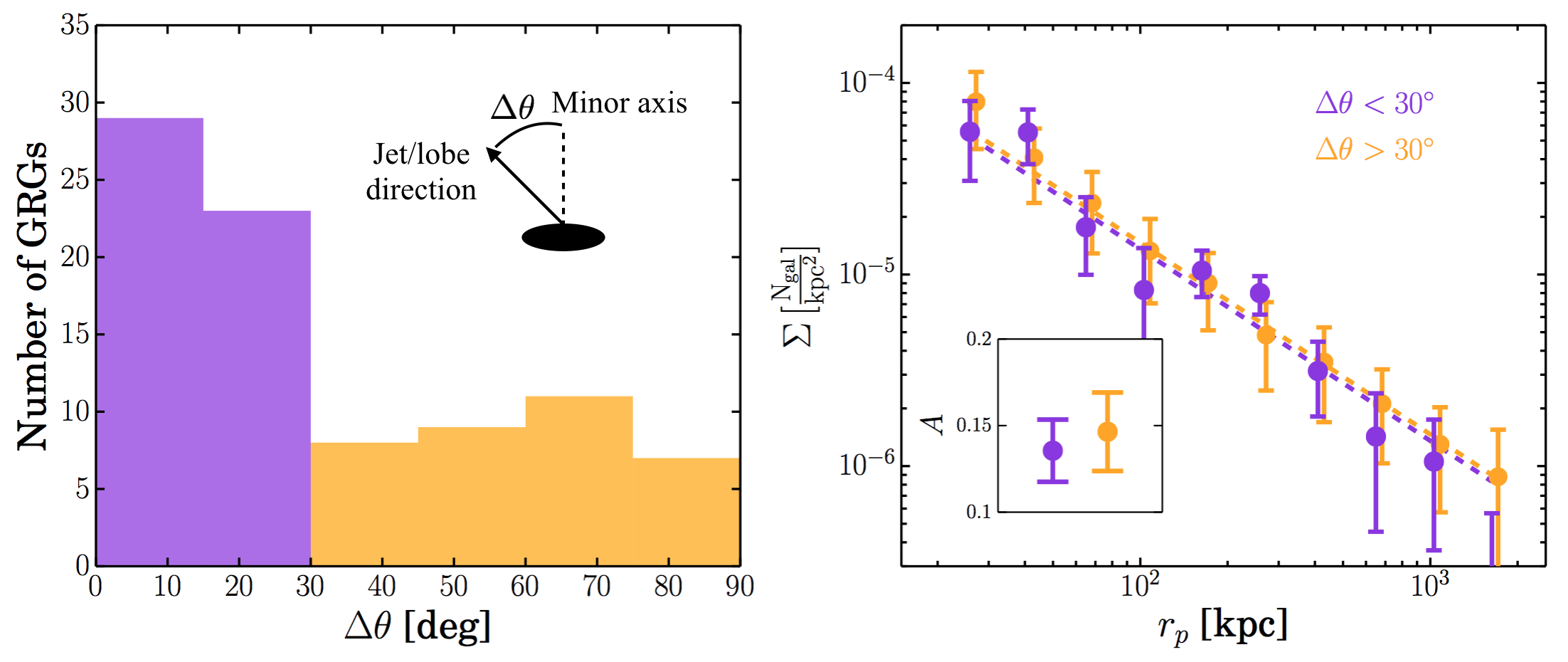}
\caption{Alignment of host galaxies and radio lobes.
\emph{Left:} Angle difference $\Delta \theta$ between the minor axes of the host galaxies and the jet/lobe directions with small (large) values indicating that jets/lobes tend be found along the minor (major) axes of the host galaxies.
\emph{Right:} Radial distribution of satellites as a function of angle difference. The distribution includes satellite galaxies with $M_{*}>10^{10.3} \, M_{\odot}$ and GRGs at $0.1<z<0.7$. The purple and orange data points indicate systems with $\Delta \theta<30^{\circ}$ and $\Delta \theta>30^{\circ}$ respectively. The best-fit A parameter values are shown in the inset.}

\label{fig:angle}
\end{figure*}
Additionally to the sizes of GRGs, we also measure the satellite distribution as a function of the azimuthal angle of galaxy locations relative to the radio lobe directions. The azimuthal angle, $\phi$, defined as the angle between the location of a galaxy and the axis of radio lobes, is illustrated by the cartoon in the right panel of Figure~\ref{fig:size_az}. We have measured the radio lobe axes of GRGs manually. 
The right panel of Figure~\ref{fig:size_az} shows the 
radial distribution of satellites for two azimuthal angle bins with $|\phi|<45^{\circ}$ for galaxies located towards the lobe directions and $|\phi|>45^{\circ}$ for galaxies located perpendicular to the lobe directions.   
Again, the two measurements are consistent with each other. 
This result demonstrates that the radio lobes of GRGs do not occur in preferential directions 
relative to the local galaxy environments. 

%--------------
%-------------------------------------

Finally, we measure the alignments between the radio lobes and the minor axis of the host galaxies. To do so, we (1) select host galaxies only with morphology best fitted by de Vaucouleurs' profiles, i.e., elliptical galaxies, characterized by the {\it Tractor} algorithm \citep{Lang2016}, (2) use the shape measurements to obtain the position angles, and (3) compare the position angles of the minor axes of galaxies with the radio lobe directions. 
The left panel of Figure~\ref{fig:angle} shows the distribution of angle differences with $\Delta \theta=0^{\circ}$ indicating that the radio lobe direction is the same as the direction of host galaxy minor axis and $\Delta \theta=90^{\circ}$ indicating that the two are completely perpendicular to each other. The cartoon in the panel illustrates $\Delta \theta$.

We find that $>50\%$ of GRGs have radio jet/lobe directions within $30^{\circ}$ of the minor axes of the host galaxies. This distribution is $\sim 3\sigma$ (p-value $\simeq 0.004$) away from a random distribution based on K-S test for two samples, suggesting that a majority of the radio jets and lobes of GRGs preferentially escapes along the minor axes of their host galaxies. This trend is consistent with normal elliptical galaxies with extended radio emission as shown in \citet{Battye2009}.
The right panel of Figure~\ref{fig:angle} shows the satellite distributions of the two GRG sub-samples with one having $\Delta \theta<30^{\circ}$ and the other having $\Delta \theta>30^{\circ}$. No correlation between the alignments of radio lobes and the axes of host galaxies and the satellite properties is detected. 

\begin{figure*}
\center
\includegraphics[width=0.95\textwidth]{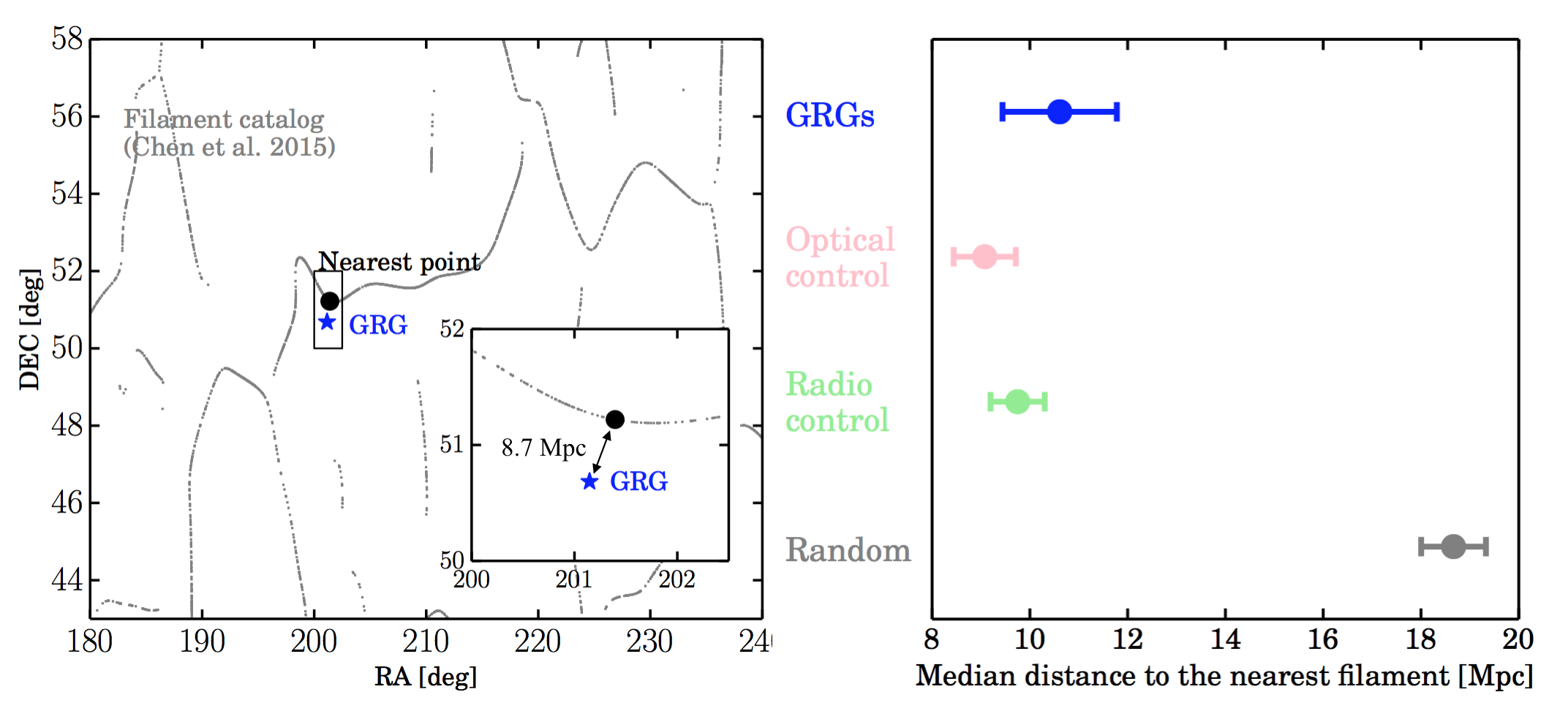}
\caption{\emph{Left:} Illustration of the filament structures. The blue data point is a GRG and the grey data points are the random positions of the filament structure provided by \citet{Chen2016catalog} catalog. The black data point indicates the nearest random point on the filament to the GRG. \emph{Right:} Median distances between GRGs (blue), the optical control sample (red), the radio control sample (green) and random positions (grey) to the nearest filament structures.}
\label{fig:large_scale}
\end{figure*}

\subsection{The large-scale environments of GRGs}
We now address the large-scale environments of GRGs. To do so, we measure the distance between each GRG and its nearest filamentary structure from the filament catalog \citep{Chen2016catalog}. The left panel of Figure~\ref{fig:large_scale} shows an example. The blue data point shows a GRG and the grey points are the random sampling from the filaments with the redshift interval being closest to the redshift of the GRG. The black data point indicates the nearest random point from the filament. The projected physical distance between the GRG and the nearest random point is estimated. 
We perform this estimation for all the GRGs at $z<0.7$ and calculate the median distance of all the GRGs to the nearest filaments. We note that this value should not be considered as the intrinsic distance between the two entities given that the distance is based on the nearest random points on the filaments. Nevertheless, this measurement is sufficient for our purpose to make comparison between the distances of GRGs and the control samples to the nearest filaments. 

The right panel of Figure~\ref{fig:large_scale} shows the result. We find that the median distance of GRGs to the nearest filaments is approximately 10 Mpc (blue), which is consistent with the same measurements of the two control samples as shown by the red and green data points. We also perform the same calculation using random points in the footprint with the result indicated by the grey data point. As can be seen, GRGs and the control samples are significantly closer to the nearby large-scale structures than random positions, illustrating that the estimator is sensible. The consistency between the measurements of GRGs and the control samples suggests that GRGs do not preferentially live in lower density environments
relative to galaxies with similar optical properties.

\section{Discussion}
\subsection{Comparison with previous studies}
We have shown that the local and large-scale environments of GRGs are consistent with that of the control samples. This finding indicates that environments play a minor role (if any) on the physical processes giving rise to the origins of GRGs. This result is consistent with some previous results. For example, \citet{Komberg2009} performed a similar analysis with $\sim20$ GRGs at $z<0.1$ and found that the local galaxy density of GRGs is similar to the galaxy density around normal radio galaxies. A recent analysis by \citet{Dabhade2020} also shows that some GRGs are located in galaxy clusters, disfavoring the low-density environment hypothesis. 
In addition, if the IGM density plays a role in confining the sizes of GRGs, one would expect the sizes of GRGs evolve with redshift. However, such a trend is not observed \citep[e.g.,][]{Machalski2006,Kuzmicz2018}, supporting the same picture that environments are not the primary factor for the sizes of GRGs.

On the other hand, some previous studies show different results. For instance, \citet{Malarecki2015} argue that there is a correlation between GRGs and their surrounding environments in contrast to our findings. \citet{Chen2011} also show that satellite galaxies around a GRG tend to be found in the direction of radio lobes, suggesting a correlation between radio jet/lobe directions and the galaxy distribution. This finding is in contrary to our results as shown in Figure~\ref{fig:size_az}. We note that sample variance could be one possible explanation for the inconsistency given that these studies are based on relative small samples. Future, larger GRG catalogs will be used to confirm our results. 

\subsection{The origins of GRGs}
Our results suggest that environments are not the main driver for the enormous sizes of GRGs. This indirectly supports the scenario that such a radio structure is driven by the internal properties of the radio jet activity. One possible scenario is that GRGs are formed with a long lifetime of radio activity as suggested by some previous studies. For example, \citet{Hardcastle2018} shows that with a fixed gas density environment, the sizes of radio galaxies depend on the lifetime of radio activity. Using the model from \citet{Hardcastle2018}, \citet{Hardcastle2019} further show that the sizes and power of observed radio galaxies, including GRGs, can be reproduced. Their results demonstrate that the sizes of radio lobes can reach to $\sim1$ Mpc after a few hundred Myr of radio activity. While the model assumes a single phase of radio activity, it might not be realistic given that AGN activity can vary within much shorter timescales \citep[e.g.,][]{Aranzana2018}. This can be reconciled by the accumulation of multiple episodes of radio activity as indicated by previous studies \citep[e.g.,][]{Bruni2020}. 

%-------------------------------------
\begin{figure}
\center
\includegraphics[width=0.48\textwidth]{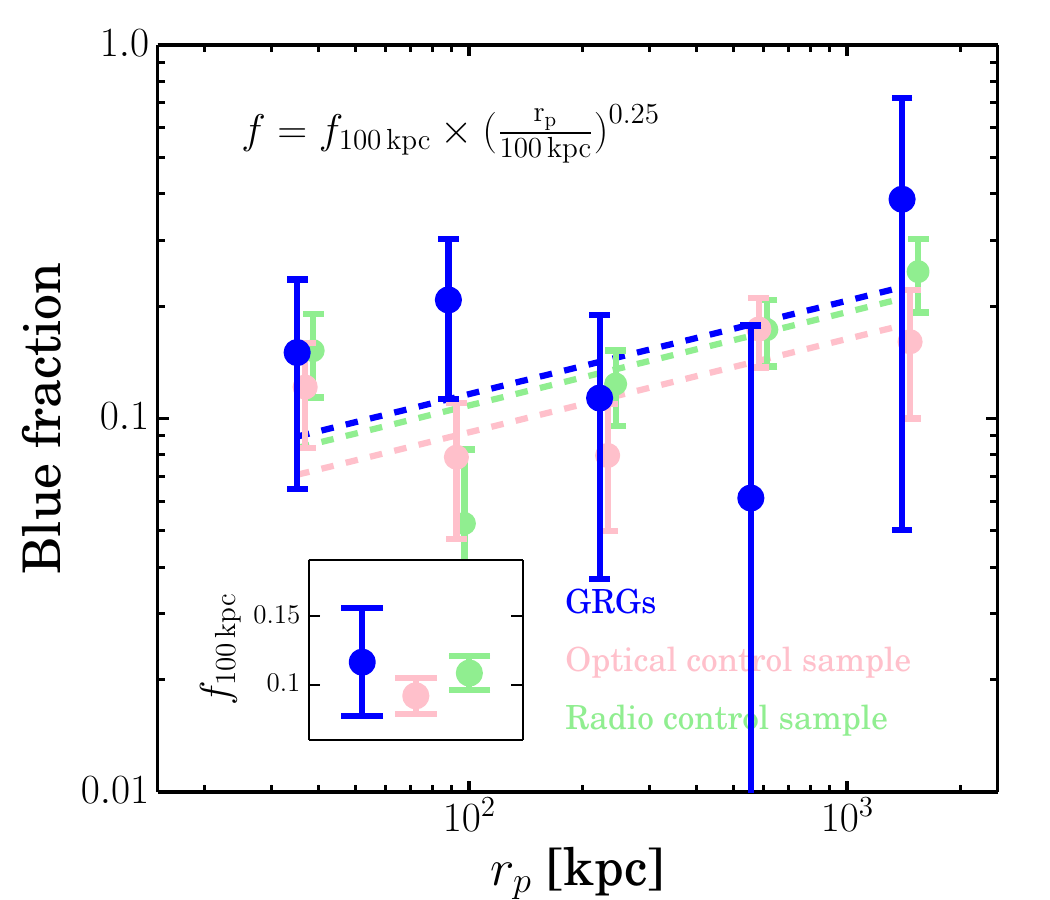}
\caption{Fraction of blue satellite galaxies with $M_{*}>10^{10.3} \, M_{\odot}$ around GRGs (blue) and the two control samples (red: optical, green: radio) with $0.1<z<0.7$. The inset shows the best-fit values of the blue fraction for the three samples at 100 kpc. No significant correlation between the fraction of blue galaxies and the radio properties is observed.}
\label{fig:blue_satellte_fraction}
\end{figure}

%\begin{figure*}
%\center
%\includegraphics[width=1\textwidth]{200817_new_blue_az_size.pdf}
%\caption{\emph{Left:} Radial distribution as a function of GRG size for blue satellite galaxies with $M_{*}>10^{9.8} \, M_{\odot}$ around GRGs with $0.1<z<0.7$. The purple (orange) data points show the surface number density of galaxies around GRGs with sizes greater (less) than 860 kpc. 
%\emph{Right:} Radial distribution as a function of azimuthal angle for blue satellite galaxies. The purple data points show the surface number density of galaxies along the radio lobe directions and the orange data points away from the radio lobe directions. The insets in two panels show the best-fit values of the radial distributions of satellites at 100 kpc. 
%From the two panels, we can conclude that the presence of enormous radio structures does not significantly affect the properties of satellite galaxies.}
%\label{fig:blue_az_size}
%\end{figure*}
\subsection{The effects of radio feedback on the properties of surrounding galaxies}
\label{sec:quenching}
While we have primarily focused on addressing the mechanisms giving rise to the enormous sizes of GRGs, we can ask the following question in the context of radio feedback - do the current radio jets of GRGs influence the satellite galaxy properties? It has been observed, for example, that the power of radio jets is sufficient to remove hot gas within their trajectory and 
thereby create cavities of hot gas \citep[e.g.,][]{Fabian2012}. 
It is possible that such activity could also remove gas 
from satellite galaxies and quench those satellites. 
However, we emphasize that our measurements are only sensitive to the effects of the current radio jets on the properties of galaxies. Episodic radio activity and the corresponding feedback processes can impact the properties of galaxies on a longer timescale that is not reflected in our measurements.

To explore radio mode feedback on quenching satellite galaxies, we first measure the fraction of blue satellite galaxies around GRGs and the two control samples. The blue satellite galaxies are selected with three observed colors, $g-z<1.7$, $g-z<2.1$, and $g-z<2.6$, for three redshift ranges $0.1<z<0.3$, $0.3<z<0.5$, and $0.5<z<0.7$ respectively. These observed color cuts are chosen to optimally separate blue and red galaxy populations based on the galaxy catalog from the PRIMUS survey \citep[][]{Coil2011, Cool2013}.

Figure~\ref{fig:blue_satellte_fraction} shows the fraction of blue satellites around GRGs and the two control samples. 
We find that the blue fraction around GRGs is consistent with 
that around the two control samples.  Furthermore, in all 
three samples the blue fraction increases with impact parameter from $\sim10\%$ at 100 kpc to $\sim30\%$ at 2 Mpc, consistent with trends observed in galaxy groups and clusters \citep[e.g.,][]{Hansen2009}. We characterize this trend with $f=f_{100\, kpc}\times \rm (r_{p}/ 100\, kpc)^{0.25}$ with the power index 0.25 being the best-fit value for the three measurements. The best-fit fraction at 100 kpc 
($f_{100\, kpc})$
is shown in the inset of the figure. This result indicates that GRGs do not preferentially have more passive satellite galaxies despite their very large radio structures. 

We also explore the effect of the current radio activity on galaxy properties by comparing the number of blue satellites as a function of GRG size and the galaxy location with respect to the radio lobes. We find that the number of blue galaxies depend on neither the sizes of GRGs nor the locations of galaxies relative to the radio lobes.
%The right panel of Figure~\ref{fig:blue_az_size} shows the radial distributions of blue satellites around GRGs ($0.1<z<0.7$) as a function of GRG size. Here we include blue satellite galaxies with $>10^{9.8} \, M_{\odot}$. The dashed lines in the figure indicate the best-fit power law with $\Sigma=A\times10^{-4}\times(r_{p}/100 \, kpc)^{0.75}$. We fix the power index to 0.75, a consistent value with the best fit values for the two measurements when the parameter is free. The inset shows the best-fit A parameter values for the two bins. The left panel shows the radial distributions of blue satellites around GRGs with two azimuthal angle bins as defined previously in Figure~\ref{fig:angle} with $|\phi|<45^{\circ}$ representing galaxies close to the radio lobes and with $|\phi|>45^{\circ}$ representing galaxies located away from the radio lobes.  
%As shown in the figure, the number of blue galaxies depend on neither the sizes of GRGs nor the locations of galaxies relative to the radio lobes. 

These results demonstrate that there is no apparent correlation between the properties of satellite galaxies and the presence of the giant radio structures, implying that 
%for satellite galaxies with $M_{*}>10^{9.8}\, M_{\odot}$, 
the current radio jets do not greatly influence satellite properties.  
This result is consistent with the result of \citet{Pace2014} who investigate the effect of typical radio AGNs on their surrounding galaxies. However, \citet{Shabala2011} show a different result. They find that galaxies located in the radio lobes of FRII-type galaxies are redder than galaxies outside of the radio lobes, offering evidence that radio jets quench star-formation in satellite galaxies. 
However, we note that \citet{Shabala2011} only considered galaxies located within radio lobes which consist of $\sim 1\%$ of all the satellite galaxies around FRII galaxies in their sample. 
In our analysis, we consider the bulk of satellite galaxies which
may explain the apparent disagreement between the two results. 
We note that with the new LoTSS dataset, one can perform the same analysis as applied in \citet{Shabala2011}. Nonetheless, such an analysis is beyond the scope of this paper.

%----
%--------------

\section{Summary}
We investigated the local and large-scale environments of GRGs by measuring the galaxy density around GRGs and their relative locations to the nearby large-scale structures.
These measurements were derived from by cross-correlating a large GRG sample detected from LoTSS, with photometric galaxies from the DESI Legacy Surveys and a filament catalog obtained from the SDSS. 
Our main findings are summarized as follows:
\begin{enumerate}
    \item We find that the properties of satellite galaxies, including number counts, magnitude and stellar mass distributions, and radial distribution, around GRGs from 20 kpc to 2 Mpc are consistent with the properties of satellite galaxies around the control samples. This result suggests that the local environments of GRGs mostly depend on the optical properties of the host galaxies. 
    
    \item We further explore the properties of satellites as a function of GRG properties and find that there is no correlation between the satellite properties and the sizes of GRGs as well as the locations of satellites relative to the radio jets/lobes.
    
    \item We estimate the locations of GRGs with respect to the nearby large-scale filament structures and find that the median distance between GRGs and the nearest filaments is consistent with the median distances between the control samples and the nearest filaments. This result indicates that GRGs do not reside in low density environments in the context of large-scale structure.  
\end{enumerate}
These results indicate that the local and large-scale environments of GRGs are mostly determined by processes within the host galaxies. No significant correlation between the existence of giant radio structures and their surrounding environments traced by stars is found. These findings suggest that the environments of GRGs play little role in the mechanisms giving rise to the giant radio structures.
In other words, our results disfavor the hypothesis that the formation of GRGs is related to their local environments \citep[e.g.,][]{Mack1998}. 

In the context of feedback, we explore the effect of the current radio activity on the properties of satellite galaxies. We find that the number of blue satellites does not depend on the sizes of GRGs nor their relative locations with respect to the radio lobes. In addition, the fraction of blue satellite galaxies around GRGs is consistent with the fraction around the control samples. These findings illustrate that the current radio activity has minimal influence on the properties of the surrounding galaxies.

In the near future, the complete LoTSS survey will detect $\sim 12,000$ GRGs across half of the sky \citep{Dabhade2020}. The combination of this large GRG sample, optical sky survey datasets, such as DESI \citep{Levi2013}, Euclid \citep{Amiaux2012}, 
and LSST \citep{Ivezic2019}, and X-ray sky survey datasets, such as eROSITA \citep{Merloni2012} will enable much more precise characterizations of the properties of the GRG host galaxies, the surrounding environments of GRGs, and the corresponding redshift evolution. 
In addition to giant radio galaxies, there are giant radio sources hosted by quasars found from the local Universe to beyond redshift 2. 
An interesting research direction is to perform IFU observations for these high-redshift giant radio quasars to map out the distribution of diffuse cool gas via Lyman alpha emission - a different tracer for the environments of high-redshift giant radio quasars \citep[e.g.,][]{Can2014}.
By doing so, we will shed new light on our understanding of not only the nature and origin of these giant radio structures but also the role they play in galaxy formation and evolution.

\acknowledgements
We thank Yi-Hao Chen for useful discussions.
TWL and JXP acknowledge support from NSF grant AST-1911140.
We also thank the LoTSS team for making the data publicly available. 
Kavli IPMU is supported by World Premier International Research Center Initiative of the Ministry of Education, Japan.

The DESI Legacy Surveys consist of three individual and complementary projects: the Dark Energy Camera Legacy Survey (DECaLS; NOAO Proposal ID $\#$ 2014B-0404; PIs: David Schlegel and Arjun Dey), the Beijing-Arizona Sky Survey (BASS; NOAO Proposal ID $\#$ 2015A-0801; PIs: Zhou Xu and Xiaohui Fan), and the Mayall z-band Legacy Survey (MzLS; NOAO Proposal ID $\#$ 2016A-0453; PI: Arjun Dey). DECaLS, BASS and MzLS together include data obtained, respectively, at the Blanco telescope, Cerro Tololo Inter-American Observatory, National Optical Astronomy Observatory (NOAO); the Bok telescope, Steward Observatory, University of Arizona; and the Mayall telescope, Kitt Peak National Observatory, NOAO. The DESI Legacy Surveys project is honored to be permitted to conduct astronomical research on Iolkam Du'ag (Kitt Peak), a mountain with particular significance to the Tohono O'odham Nation.

NOAO is operated by the Association of Universities for Research in Astronomy (AURA) under a cooperative agreement with the National Science Foundation.

This project used data obtained with the Dark Energy Camera (DECam), which was constructed by the Dark Energy Survey (DES) collaboration. Funding for the DES Projects has been provided by the U.S. Department of Energy, the U.S. National Science Foundation, the Ministry of Science and Education of Spain, the Science and Technology Facilities Council of the United Kingdom, the Higher Education Funding Council for England, the National Center for Supercomputing Applications at the University of Illinois at Urbana-Champaign, the Kavli Institute of Cosmological Physics at the University of Chicago, Center for Cosmology and Astro-Particle Physics at the Ohio State University, the Mitchell Institute for Fundamental Physics and Astronomy at Texas A$\&$M University, Financiadora de Estudos e Projetos, Fundacao Carlos Chagas Filho de Amparo, Financiadora de Estudos e Projetos, Fundacao Carlos Chagas Filho de Amparo a Pesquisa do Estado do Rio de Janeiro, Conselho Nacional de Desenvolvimento Cientifico e Tecnologico and the Ministerio da Ciencia, Tecnologia e Inovacao, the Deutsche Forschungsgemeinschaft and the Collaborating Institutions in the Dark Energy Survey. The Collaborating Institutions are Argonne National Laboratory, the University of California at Santa Cruz, the University of Cambridge, Centro de Investigaciones Energeticas, Medioambientales y Tecnologicas-Madrid, the University of Chicago, University College London, the DES-Brazil Consortium, the University of Edinburgh, the Eidgenossische Technische Hochschule (ETH) Zurich, Fermi National Accelerator Laboratory, the University of Illinois at Urbana-Champaign, the Institut de Ciencies de l'Espai (IEEC/CSIC), the Institut de Fisica d'Altes Energies, Lawrence Berkeley National Laboratory, the Ludwig-Maximilians Universitat Munchen and the associated Excellence Cluster Universe, the University of Michigan, the National Optical Astronomy Observatory, the University of Nottingham, the Ohio State University, the University of Pennsylvania, the University of Portsmouth, SLAC National Accelerator Laboratory, Stanford University, the University of Sussex, and Texas A$\&$M University.

BASS is a key project of the Telescope Access Program (TAP), which has been funded by the National Astronomical Observatories of China, the Chinese Academy of Sciences (the Strategic Priority Research Program "The Emergence of Cosmological Structures" Grant $\#$ XDB09000000), and the Special Fund for Astronomy from the Ministry of Finance. The BASS is also supported by the External Cooperation Program of Chinese Academy of Sciences (Grant $\#$ 114A11KYSB20160057), and Chinese National Natural Science Foundation (Grant $\#$ 11433005).

The Legacy Survey team makes use of data products from the Near-Earth Object Wide-field Infrared Survey Explorer (NEOWISE), which is a project of the Jet Propulsion Laboratory/California Institute of Technology. NEOWISE is funded by the National Aeronautics and Space Administration.

The DESI Legacy Surveys imaging of the DESI footprint is supported by the Director, Office of Science, Office of High Energy Physics of the U.S. Department of Energy under Contract No. DE-AC02-05CH1123, by the National Energy Research Scientific Computing Center, a DOE Office of Science User Facility under the same contract; and by the U.S. National Science Foundation, Division of Astronomical Sciences under Contract No. AST-0950945 to NOAO.

\section{Data availability}
The data underlying this article will be available at \url{https://people.ucsc.edu/~tlan3/research/GRGs/}.

{}
\end{CJK*}
\end{document}